\documentclass{iau}
\usepackage{graphicx}
\usepackage{amsmath}

\newcommand{\corot} {\emph{CoRoT}}
\newcommand{\kepler} {\emph{Kepler}}
\newcommand{\soho} {\emph{SoHO}}

\newcommand{\ds} {$\delta$~Scuti}
\newcommand{\gd} {$\gamma$~Dor}
\newcommand{\gds} {$\gamma$~Dor stars}

\newcommand{\conn} {{\cal C}_n}
\newcommand{\enn} {{\cal D}_n}

\newcommand{\apj} {ApJ}

\newcommand{\eqn} [1] {
\begin{equation}
#1
\end{equation}}

\title{On the necessity of a new interpretation of the stellar light curves}

\author[J. Pascual-Granado, R. Garrido \& J.C. Su{\'a}rez]   
{J. Pascual-Granado, R. Garrido, \and J.C. Su{\'a}rez}

\affiliation{Instituto de Astrof\'isica de Andaluc\'ia (CSIC), Glorieta de la Astronom\'ia s/n 18008, Granada, Spain. 
email: {\tt javier@iaa.es} }

\pubyear{2014}
\volume{301}  
\pagerange{1--4}
\jname{Precision Asteroseismology}
\editors{J.A. Guzik, W.J. Chaplin, G. Handler \& A. Pigulski, eds.}
\begin{document}

\maketitle

\begin{abstract}
The power of asteroseismology relies on the ability to infer the stellar structure from the unambiguous frequency identification of the corresponding pulsation mode. Hence, the use of a Fourier transform is in the basis of asteroseismic
studies. Nevertheless, the difficulties with the interpretation of the frequencies found in many stars lead us to reconsider whether Fourier analysis is the most appropriate technique to identify pulsation modes. We have found that the data, usually analyzed using Fourier techniques, present a non-analyticity originating from the lack of connectivity of the underlying function describing the physical phenomena. Therefore, the conditions for the Fourier series to converge are not fulfilled. In the light of these results, we examine in this talk some stellar light curves from different asteroseismology space missions (\corot, \kepler\ and \soho) in
which the interpretation of the data in terms of Fourier frequencies becomes difficult. We emphasize the necessity of a new interpretation of the stellar light curves in order to identify the correct frequencies of the pulsation modes.

\end{abstract}

\firstsection 
              
\section{Overview}
Asteroseismology has been very fruitful since the launch of space satellites but some old problems remain unsolved: the non-detection of the solar g-modes in the solar spectrum is a good example. Another persistent problem is the poor modelling of outer layers of the Sun requiring an ad-hoc correction now extended to solar-like stars by \cite{Kjeldsen}. No physical justification for this correction has ever been given. The presence of constant stars within
the instability strip of \ds\ and \gds\ (Guzik et al., these proceedings) is another puzzle. 
More importantly,  the range and number of frequencies detected in many pulsating stars observed by satellites are not yet understood. For instance, the observed instability range of the 422 frequencies of the \ds\ star HD\,174936 could not be reproduced by any model (\cite[Garc\'ia Hern\'andez et al.~2009]{AGH}). For HD\,50844 (\cite[Poretti et al.~2009]{Poretti}), prewhitening of more than thousand frequencies yielded a bushy structure in its residuals, very different from the expected white noise.
HD\,50870 is another example showing the same phenomenon  (\cite[Mantegazza et al.~2012]{Mantegazza}). Additionally, the analysis of the \gd\ star HD\,49434 by \cite{Chapellier} showed that the amplitudes of new frequency components decrease exponentially as the number of fitted frequencies increases linearly. All the mentioned cases come from the observations made by the \corot\ satellite (\cite[Auvergne et al.~2009]{Auvergne}), but recently some paradigmatic cases have appeared in the \kepler's sample of pulsating stars (\cite[Gilliland et al.~2010]{Gilliland}): high frequencies in the oscillation spectrum of the  \ds\ star  KIC 4840675 of  unknown nature (\cite[Balona et al.~2012]{Balona}), or  KIC 8677585, the only roAp star in which low frequencies are clearly detected (\cite[Balona et al.~2013]{Balona13}). 

All the efforts made to explain frequency spectra have been always on the side of the theoretical modelling,  never questioning the analysis, which is supposed to be consistent but never has been tested.  We put the focus here on this possibility.

\section{Methods}
Normally we use a DFT to calculate the periodogram (\cite[Scargle 1982]{Scargle}) and perform a frequency detection based on a Fisher test. The consistency of this operation is due to Parseval's theorem (\cite[Kaplan 1992]{Kaplan}). The DFT and its inverse are approximations to an expansion in Fourier series. When the Fourier series does not converge the function has no expansion and the DFT has no sense. 
Therefore it is very important to check if the underlying function of a discrete series can be expanded in Fourier series and this is only guaranteed when the function is analytic (\cite[van Dijk 2009]{Dijk}). Then, before applying any further analysis we should test the analyticity characterized here through the differentiability. 	 

Now, to perform the differentiability analysis of the underlying function our approach is to study how compactly connected the discrete data are through a property called connectivity. We remark here that what we are studying is the differentiability of the continuous function underlying the discrete data and not the discrete data themselves. 

The connectivity $\conn$ of a data point $x_n$ is defined as

\begin{align}
C_n & = \epsilon^f_n - \epsilon^b_n \qquad \text{with $\epsilon^f_n$, $\epsilon^b_n$ defined as} \\
\epsilon^f_n & = x^f_n - x_n \nonumber \\
\epsilon^b_n & = x^b_n - x_n \nonumber
\end{align}
         
          
where $x^f_n$ and $x^b_n$ are forward and backward extrapolations respectively made from the data bracketing a datapoint $x_n$. From these equations the numerical approximation of the point derivative $\enn$ at $x_n$ can be expressed through the connectivity as:
\eqn{\enn = \frac{\conn + x_{n+1}-x_{n-1}}{2 \Delta t}\label{eq:def_enn}}
which reduces to the typical point derivative for discrete data when $\conn = 0$. When connectivities are not zero but independent randomly distributed values, the derivative is still well-defined. In that case, the connectivities can be considered simply as deviations from the derivative at this point. Otherwise, the derivability condition is not fulfilled.

As it is defined the connectivity resembles the non-differentiability coefficient introduced by Wiener (\cite[Wiener 1923]{Wiener}) but involving extrapolations in our case.

To calculate those extrapolations we need a model that can approximate arbitrarily well any continuous function. The Stone-Weierstrass theorem (\cite[Royden 1988]{Royden}) states that a function uniformly continuous in a closed interval can be approximated arbitrarily well by a polynomial of degree $n$. Therefore, provided a sufficient amount of data, a spline function (\cite[de Boor 1978]{splines}) should provide a good extrapolation for the datapoint - the residuals form an independent random sequence, i.e. white noise. We also make use of an ARMA (autoregressive moving average) approximation (\cite[Box \& Jenkins 1976]{Box}) which is capable of fitting non-analytic functions.  Both results are compared. 

\section{Results}
We have calculated the connectivities in the following cases: 
\begin{itemize}
 \item The light curve of a \ds\ star observed by \corot, and the best analytic model that can be obtained from the Fourier analysis of its light curve.
 \item A hybrid star observed by \kepler
 \item Radial velocities of the Sun observed by the GOLF instrument onboard \soho\
\end{itemize}

The first case is the \ds\ HD\,174936, one of the enigmatic cases having a range of excited frequencies not predicted by any model. An analytic model for this star has been constructed using the 422 frequencies detected using Fourier techniques in \cite{AGH}. It is expected that both time series show the same properties. 

\begin{figure*}	
   \includegraphics[width=2.7in]{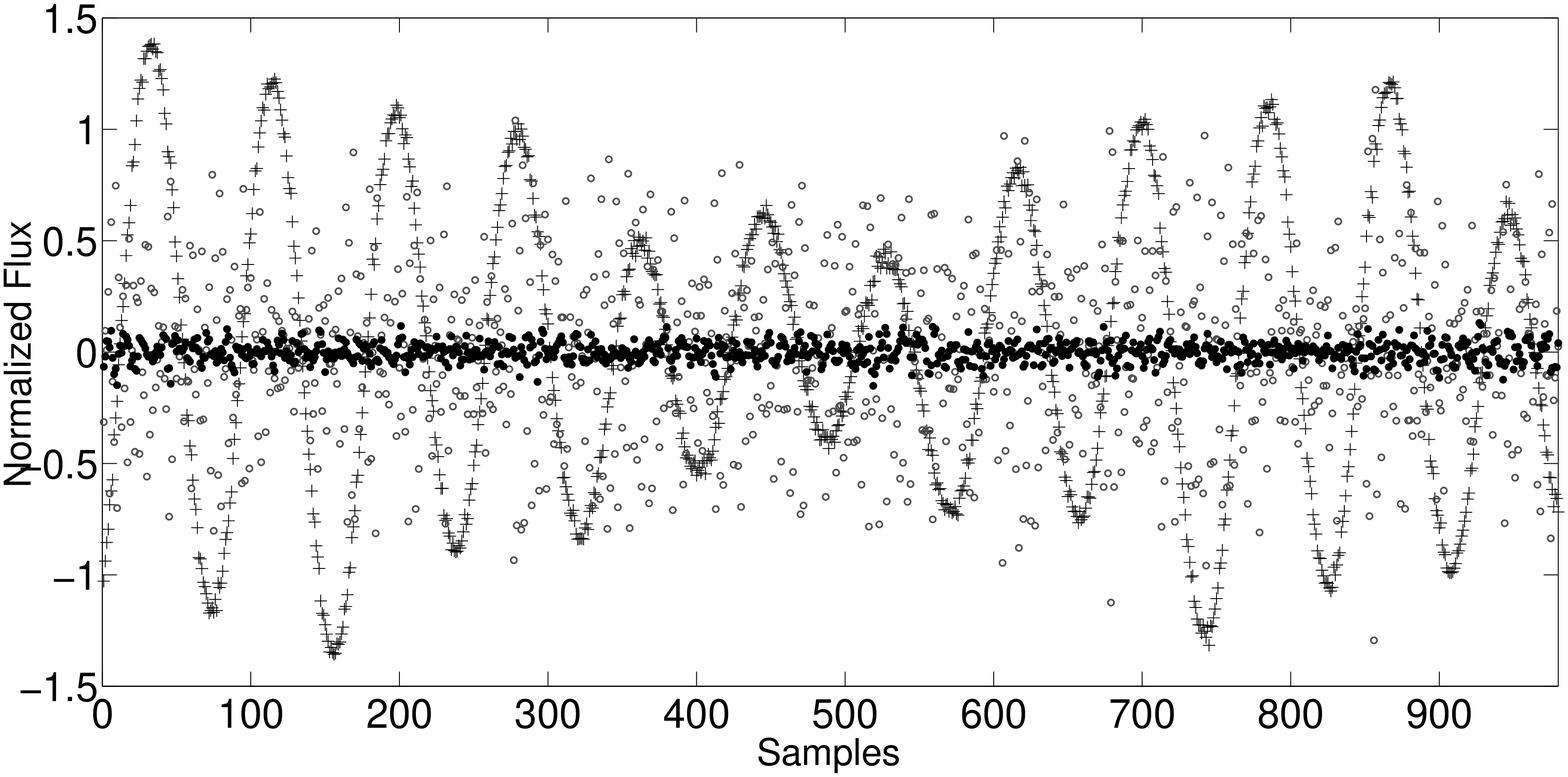}
   \includegraphics[width=2.7in]{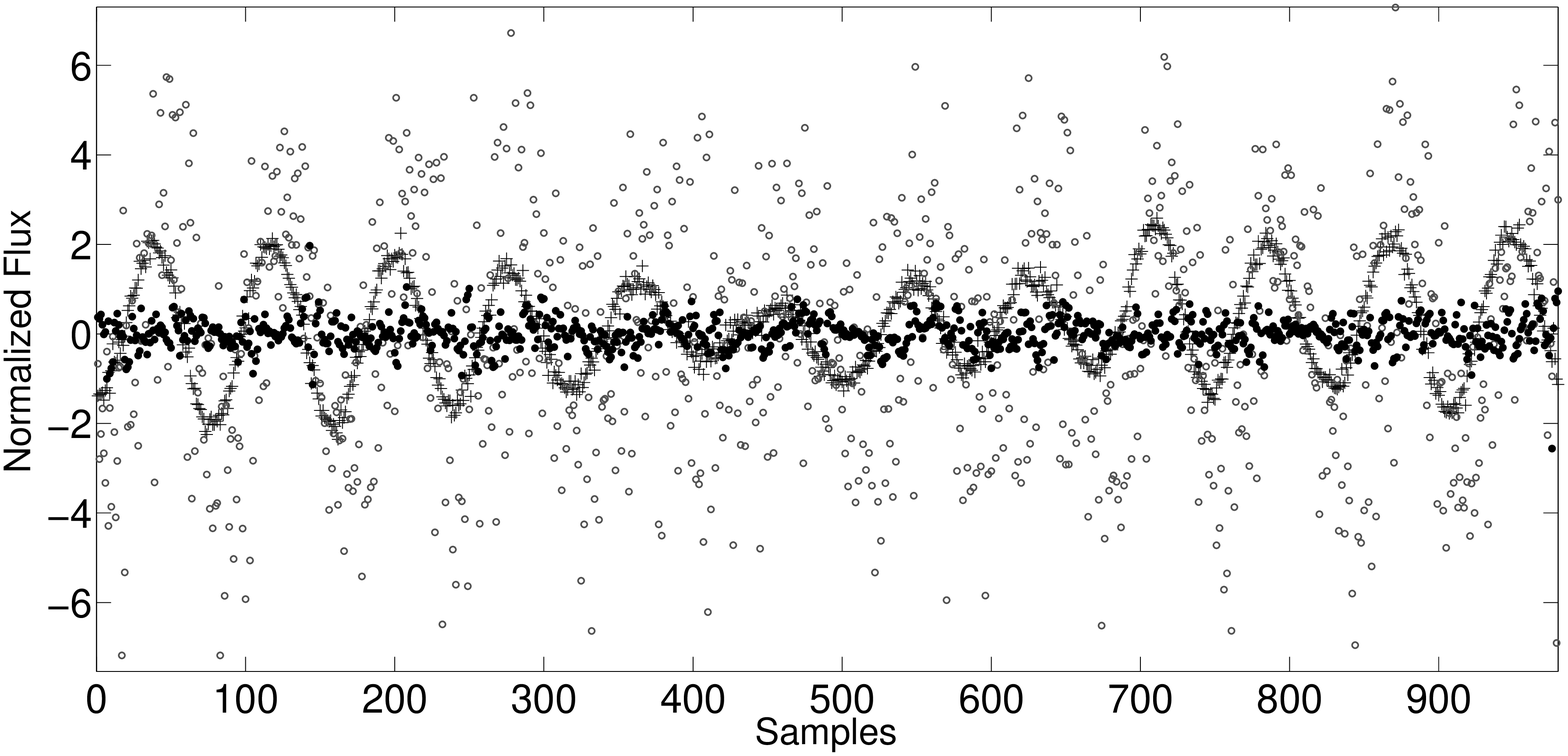}
   \caption{Right: \corot\ data for the \ds\ star HD\,174936. Left: the corresponding analytic model. Connectivities - ARMA (points), splines (circles), and the original light curve (crosses). Note the different scaling of the panels.}    
   \label{fig1}
\end{figure*}

In Fig.~\ref{fig1} we show the connectivities calculated with both approaches: splines (analytic fitting) and ARMA (non-analytic fitting). We emphasize here that the points should show an independent random distribution in case of a well-behaved underlying function that could be interpreted in terms of Fourier frequencies. However, when considering only the spline connectivities,  while the analytic model shows that kind of distribution, the connectivities of the \corot\ data are completely correlated. Two conclusions come to mind from this first analysis: 

\begin{itemize}
 \item \corot\ data for the \ds\ HD\,174936 do not originate from an analytic function.
 \item The analytic model built from Fourier frequencies does not represent the same function as \corot\ data.
\end{itemize}

When we consider ARMA connectivities, in contrast to the analytic approach given by the splines, these connectivities show a distribution typical of a white noise in both panels of Fig.~\ref{fig1}, giving more weight to the previous conclusions.

\begin{figure}
    \centering
    \resizebox{8cm}{!}{\includegraphics{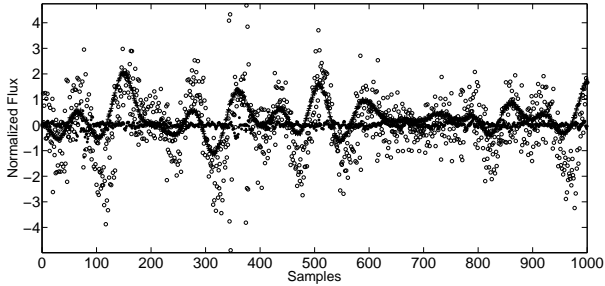}}
   \caption{\kepler\ data for the hybrid star KIC 006187665. Connectivities - ARMA (points), spline (circles), and the original light curve (crosses).}
    \label{fig2}
\end{figure}

The second light curve studied here is a hybrid pulsating star observed by \kepler. The connectivities calculated with splines (Fig.~\ref{fig2}) are correlated, whereas those calculated with the non-analytic approach are not. Since we are analyzing data from another instrument, no instrumental effect can be involved in the non-differentiability of the function, but rather an intrinsic effect of the light curves.

\begin{figure}	
   \includegraphics[width=2.7in]{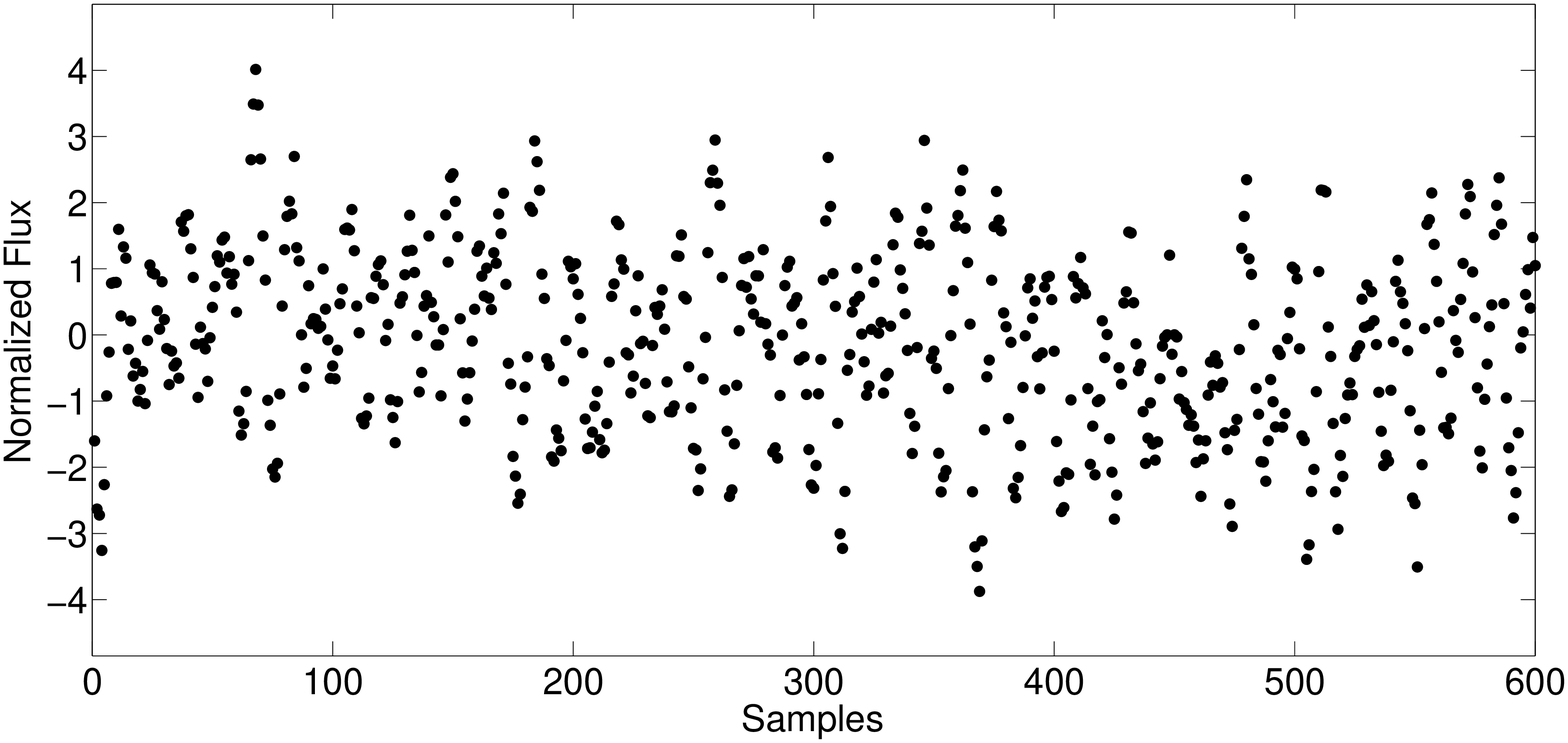}
   \includegraphics[width=2.7in]{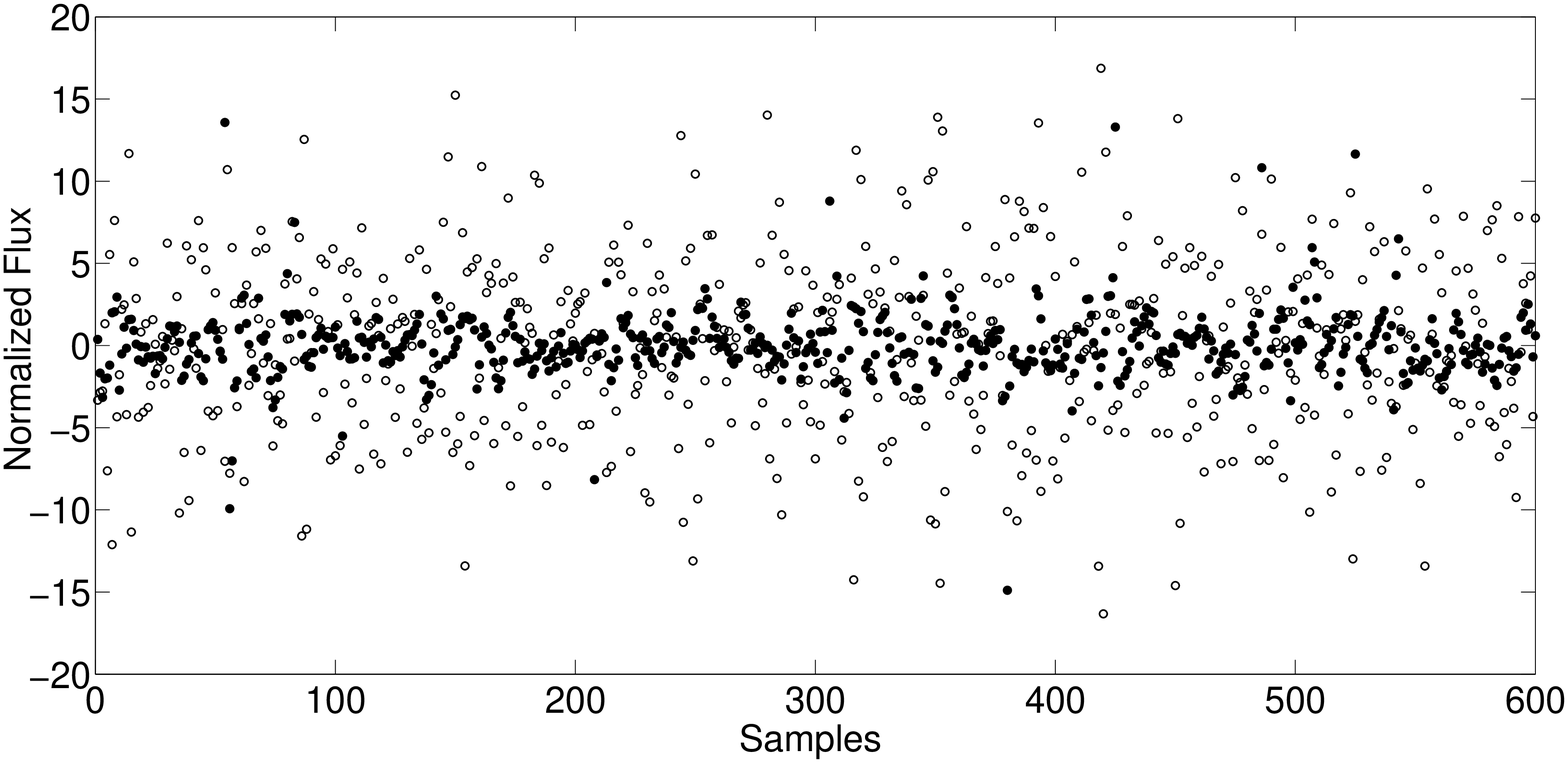}
   \caption{Left: original \soho/GOLF data. Right: connectivities - ARMA (points) and splines (circles)}    
   \label{fig3}
\end{figure}

Finally, the results corresponding to the radial velocities of the Sun taken by the GOLF instrument (\cite[Garc\'ia et al.~2005]{Garcia}) are shown in Fig.~\ref{fig3}. In this case the connectivities calculated using splines present again a higher dispersion than the ARMA ones. Connectivities are also correlated with the original time series, confirming the non-differentiability of the function in this case too. 

In summary, the non-differentiability is ubiquitous in the stellar light curves.

\section{Conclusions}
The tests carried out here show that the underlying function describing the light variations of some pulsating stars is non-analytic. This implies that the conditions for the Parseval theorem might be not satisfied. Therefore, it is not guaranteed that the signal can be represented by a Fourier series and thereby the periodogram could be not a consistent estimator of the frequency content of the underlying function. We have demonstrated that this phenomenon is neither due to an instrumental effect nor caused by the type of measurement. It does not depend on the intrinsic variability type either.
The non-analyticity of the function underlying the light curves is a fine structure that could be the origin of some puzzling features like the flickering observed in classical Cepheids (Evans et al., these proceedings).  We conclude that the standard periodogram interpretation of the time series of pulsating stars should be accordingly revised. Work is in progress to investigate the origin of this inconsistency in the harmonic analysis.

\end{document}